\documentclass[prd, 
  twocolumn,
  superscriptaddress,tightenlines,nofootinbib,
  eqsecnum,
  showpacs]{revtex4}

\usepackage{amsmath}
\usepackage{amsfonts}
\usepackage{amssymb}
\usepackage{bm}
\usepackage{hyperref}
\usepackage{mathrsfs}
\usepackage{graphicx}

\usepackage{ulem}
\usepackage[usenames]{color}

\definecolor{darkgreen}{rgb}{0,0.5,0}

\hypersetup{
    bookmarks=true,         
    unicode=false,          
    pdftoolbar=true,        
    pdfmenubar=true,        
    pdffitwindow=false,     
    pdfstartview={FitH},    
    pdftitle={My title},    
    pdfauthor={Author},     
    pdfsubject={Subject},   
    pdfcreator={Creator},   
    pdfproducer={Producer}, 
    pdfkeywords={keyword1} {key2} {key3}, 
    pdfnewwindow=true,      
    colorlinks=true,       
    linkcolor=red,          
    citecolor=cyan,        
    filecolor=magenta,      
    urlcolor=darkgreen,           
    linktocpage=true
}

\allowdisplaybreaks

\DeclareSymbolFontAlphabet{\mathrsfs}{rsfs}
\DeclareMathAlphabet{\mathcal}{OMS}{cmsy}{m}{n}

\newcommand{\ud}{\mathrm{d}}
 
\newcommand{\ue}{\mathrm{e}} 
\newcommand{\beq}{\begin{equation}}
\newcommand{\eeq}{\end{equation}} 
\newcommand\calO{{\mathcal{O}}}

\newcounter{theorem} 

\begin{document}

\title{Ambiguity-Free Completion of the Equations of Motion of\\Compact Binary
  Systems at the Fourth Post-Newtonian Order}

\author{Tanguy Marchand}\email{tanguy.marchand@iap.fr}
\affiliation{$\mathcal{G}\mathbb{R}\varepsilon{\mathbb{C}}\mathcal{O}$,
  Institut d'Astrophysique de Paris,\\ UMR 7095, CNRS, Sorbonne
  Universit{\'e}s \& UPMC Univ Paris 6,\\ 98\textsuperscript{bis} boulevard
  Arago, 75014 Paris, France}
\affiliation{Laboratoire APC -- Astroparticule et Cosmologie, \\
  Universit{\'e} Paris Diderot Paris 7, 75013 Paris, France}

\author{Laura Bernard}\email{laura.bernard@tecnico.ulisboa.pt}
\affiliation{CENTRA, Departamento de F\'{\i}sica, Instituto Superior
  T{\'e}cnico -- IST, Universidade de Lisboa -- UL, Avenida Rovisco Pais 1,
  1049 Lisboa, Portugal}

\author{Luc Blanchet}\email{luc.blanchet@iap.fr}
\affiliation{$\mathcal{G}\mathbb{R}\varepsilon{\mathbb{C}}\mathcal{O}$,
  Institut d'Astrophysique de Paris,\\ UMR 7095, CNRS, Sorbonne
  Universit{\'e}s \& UPMC Univ Paris 6,\\ 98\textsuperscript{bis} boulevard
  Arago, 75014 Paris, France}

\author{Guillaume Faye}\email{guillaume.faye@iap.fr}
\affiliation{$\mathcal{G}\mathbb{R}\varepsilon{\mathbb{C}}\mathcal{O}$,
  Institut d'Astrophysique de Paris,\\ UMR 7095, CNRS, Sorbonne
  Universit{\'e}s \& UPMC Univ Paris 6,\\ 98\textsuperscript{bis} boulevard
  Arago, 75014 Paris, France}

\date{\today}

\begin{abstract} 
  We present the first complete (i.e., ambiguity-free) derivation of the
  equations of motion of two nonspinning compact objects up to the 4PN order,
  based on the Fokker action of point particles in harmonic coordinates. The
  last ambiguity parameter is determined from first principle, by resorting to
  a matching between the near zone and far zone fields, and a consistent
  computation of the 4PN tail effect in $d$ dimensions. Dimensional
  regularization is used throughout for treating IR divergences appearing at
  4PN order, as well as UV divergences due to the modeling of the compact objects
  as point particles.
\end{abstract}


\maketitle

\section{Introduction} 
\label{sec:intro}

The recent detection of gravitational waves (GW) generated by inspiralling and
merging black-hole or neutron-star binaries~\cite{GW150914, GW170817}
highlights the importance of the problems of motion and radiation for systems
of compact objects in general relativity. Analytical relativity, based on the
post-Newtonian (PN) approximation, i.e., a formal expansion when the speed of
light $c\to \infty$, plays a key role in the development of high-accuracy GW
templates to be used in the signal analysis of detectors. The templates are
cross-correlated with the detector's output, and the correlation builds up
when a good match occurs between a particular template and the real
signal~\cite{Th300, BuonSathya15}. This technique is highly sensitive to the
phase evolution of the signal, which, in PN templates of compact binary
coalescence, is computed from the energy balance between the decay of the
binary's energy and the GW flux. For isolated binary systems, the orbit will have circularized by
radiation reaction at the time when the signal enters the detectors' bandwidth,
so we expect that for the current generation of detectors, there is no need to invoke the balance of orbital angular momentum.

For low mass compact binaries, such as double neutron star
systems~\cite{GW170817}, the detectors are mostly sensitive to the inspiral
phase prior to the final coalescence; in that case the currently known analytical PN
templates are accurate enough for detection (at least for moderate spins). For
higher masses, like in black-hole binary systems, one must somehow connect the
PN templates to the numerical relativity (NR) results describing the final
merger and ringdown phases. The hybrid inspiral-merger-ringdown (IMR)
waveforms~\cite{Ajith11} are constructed by matching the PN and NR waveforms
in an overlapping time interval; the effective-one-body (EOB)
waveforms~\cite{BuonD99} are based on resummation techniques extending the
domain of validity of the PN approximation. The IMR and EOB waveforms
constitute key techniques in the data analysis (both on-line and off-line) of
the recent black-hole events~\cite{GW150914}.

The two basic ingredients in the theoretical PN analysis correspond to the two
sides of the energy balance equation obeyed by the binary's orbital frequency
and phase. The GW flux on the right-hand side is obtained by solving the wave
generation problem; the state-of-the-art is the 3.5PN approximation beyond the
quadrupole formula (i.e., formal order $\sim c^{-7}$; see~\cite{Bliving14} for
a review), the 4.5PN coefficient being also known~\cite{MBF16}. The energy
function on the left-hand side follows from the conservative dynamics or
equations of motion; after one century of works on the problem of motion (see
for instance~\cite{LD17, EIH, Fock, CE70, OO74a, DD81b, Dhouches, S85, Will, BFP98} and
references therein) and the completion of the 3PN dynamics~\cite{JaraS98, BFeom, DJSequiv, ABF01, DJSdim, BI03CM, BDE04}, the state-of-the-art is the
4PN approximation beyond the Newtonian force.

Calculations at the 4PN order have been undertaken by means of three methods:
(i) The Arnowitt-Deser-Misner (ADM) Hamiltonian formalism~\cite{JaraS12,
  JaraS13, DJS14, DJS16}, which led to complete results but for the appearance
of one ``ambiguity'' parameter; (ii) The Fokker Lagrangian in harmonic
coordinates~\cite{BBBFMa, BBBFMb, BBBFMc, BBFM17}, which is complete at the
exception, until recently, of one equivalent ambiguity parameter;\footnote{Two
  ambiguity parameters were introduced in Ref.~\cite{BBBFMb}. In a first
  version of Ref.~\cite{BBBFMc}, one combinaison of these ambiguity parameters
  could be determined, but an incomplete implementation of the
  $\varepsilon\eta$ regularization (see below) did not permit to conclude on
  the ``last'' ambiguity parameter. We have updated the work~\cite{BBBFMc} to
  take into account the new results presented in the present article.} (iii)
The effective field theory (EFT)~\cite{FS4PN, FStail, GLPR16, FMSS16}, which
yielded partial results up to now (the terms $\propto G^4$ still being
uncomputed) and is expected to be free of ambiguities~\cite{PR17}.

The ambiguity parameters in the ADM formalism and in the Fokker action have
been computed by resorting to perturbative gravitational self-force (GSF)
determinations of the so-called redshift variable~\cite{Det08, BarackS11}. An
analytic GSF calculation provided the 4PN coefficient in the
redshift~\cite{BiniD13}; then, the first law of compact binary
mechanics~\cite{LBW12, L15, BL17} enabled one to deduce the corresponding 4PN
coefficients in the conserved energy and periastron advance for circular
orbits, in the small mass-ratio limit, which was sufficient to fix the
ambiguities. The final result for the 4PN Fokker Lagrangian is given in Sec.~V
of~\cite{BBBFMa} with some $G^4$ terms corrected in the Appendix~A
of~\cite{BBBFMb}.\footnote{See also Ref.~\cite{BBFM17} for the Lagrangian and
  equations of motion in the frame of the center of mass and for a
  recapitulation of our result.} It is fully equivalent to the final result of
the ADM Hamiltonian given in the Appendix~A of~\cite{DJS14}.

In this article --- a companion paper of Ref.~\cite{BBBFMc} --- we detail the
resolution of the important issue of the remaining (``last'') ambiguity
parameter in the 4PN Lagrangian~\cite{BBBFMa, BBBFMb, BBBFMc}. The ambiguity
is due to the presence of infrared (IR) divergences in the Fokker action,
which are in turn associated with GW tails propagating at infinity. The tails
are secondary nonlinear waves caused by the backscattering of linear waves
onto the space-time curvature generated by the total mass of the source. As we
shall see, the solution of the problem of ambiguities lies in performing the
proper matching between the near-zone field described by the PN approximation
and the far-zone radiation field. As a result of the matching, a contribution
due to tails arises precisely at the 4PN order~\cite{BD88, B93} in the
particle's action and the conservative dynamics. Due to this tail effect, the
dynamics is nonlocal in time; this entails subtleties in the derivation of
the invariants of motion and periastron advance, which have been dealt with in
Refs.~\cite{DJS14, DJS16, BBBFMb, BL17}.

Another crucial ingredient in our approach, as well as in the EFT, is
dimensional regularization, as it cures both IR divergences and concomitant
ultra-violet (UV) divergences due to the point-particles model adopted to
describe the compact objects. Dimensional regularization was introduced as a
mean to preserve the gauge invariance of quantum gauge field
theories~\cite{tHooft, Bollini, Breitenlohner}. Here, we use it in the problem
of classical interaction of point masses, as a way to preserve the
diffeomorphism invariance of general relativity~\cite{DJSdim, BDE04}. We argue
that dimensional regularization is the only known method to successfully solve
the problem at the 4PN order.

\section{Overview of the calculation} 
\label{sec:overview}

We start from the complete gravitation-plus-matter action $S=S_g+S_m$, where
the gravitational (Einstein-Hilbert) part $S_g$ is written in the
Landau-Lifshitz form with the usual harmonic gauge-fixing term, and where
$S_{m}$ is the matter part appropriate for two point particles without spin
nor internal structure [see Eqs.~(2.1)--(2.2) in Ref.~\cite{BBBFMa}]. The
gauge-fixed Einstein field equations (GFEE) deriving from $S$ read
\begin{equation}\label{GFEE}
\Box h^{\mu\nu} = \frac{16\pi G}{c^4}\tau^{\mu\nu}\,,
\end{equation}
where $\Box$ is the flat d'Alembertian operator and where
\begin{equation}\label{taumunu}
\tau^{\mu\nu} = \vert g\vert T^{\mu\nu} + \frac{c^4}{16\pi G}
\Lambda^{\mu\nu}\,.
\end{equation}
The field variable
$h^{\mu\nu} = \vert g\vert^{1/2} g^{\mu\nu} - \eta^{\mu\nu}$ is the gothic
metric deviation from the (inverse) Minkowski metric $\eta^{\mu\nu}$, with
$g^{\mu\nu}$ standing for the inverse metric and $g$ for the metric
determinant, while $T^{\mu\nu}$ is the stress-energy tensor of the particles
and $\Lambda^{\mu\nu}$ the nonlinear gravitational source term, at least
quadratic in $h^{\mu\nu}$ or its space-time derivatives. The constant $G$ is
related to the usual Newton constant $G_N$ in 3 dimensions by
$G=G_N\ell_0^{d-3}$ where $d$ is the space dimension and $\ell_0$ an arbitrary
scale.

We shall denote by $\overline{h}^{\mu\nu}$ the PN field constructed by
standard PN iteration of the GFEE~\eqref{GFEE}; such PN solution is a
functional of the particle's world-lines $\mathbf{y}_A$ (with $A=1,2$). The
Fokker action for the binary is obtained by replacing the PN solution
$\overline{h}^{\mu\nu}[\mathbf{y}_A]$ back into the original action $S$, thus
defining $S_\text{F}[\mathbf{y}_A] = S(\overline{h}[\mathbf{y}_A])$. This
action describes the purely gravitational dynamics of the compact binary
system; it is equivalent, in the ``tree-level'' approximation, to the
effective action used by the EFT approach~\cite{GR06, FSrevue}.

The PN-expanded field $\overline{h}^{\mu\nu}$ is physically valid in the near
zone of the matter system, which is of small extent with respect to the
radiation wavelength. On the other hand, the multipole expansion, denoted
$\mathcal{M}(h^{\mu\nu})$, holds all over the exterior of the system including
the far (or wave) zone. As the multipole expansion is a solution of the
GFEE~\eqref{GFEE}, it is also a functional of the particle's world-lines. Our
approach is based on the matching between the two expansions in the
overlapping region where both approximations are valid, namely the exterior
part of the near zone, which always exists for PN sources, i.e., slowly moving
and weakly stressed sources.

The matching is achieved using a variant of the general method of matched
asymptotic expansions~\cite{BuTh70, Bu71, AKKM82}. More precisely, we impose
the matching equation which states that the PN (or near-zone) expansion of the
multipolar field should be identical to the multipole (or far-zone) expansion
of the PN field:
\begin{equation}\label{matching}
\overline{\mathcal{M}(h^{\mu\nu})} =
\mathcal{M}\bigl(\overline{h}^{\mu\nu}\bigr)\,.
\end{equation}
The general solution of the GFEE satisfying the above relation is known: The
multipolar field in the exterior region is determined as a functional of the source
parameters through the explicit expressions of the multipole
moments~\cite{B98mult, BDE04}; the PN-expanded field in the near zone reads
(generalizing results from~\cite{PB02, BFN05} to $d$ dimensions)
\begin{equation}\label{PNsol}
\overline{h}^{\mu\nu} = \frac{16\pi G}{c^4}
\,\overline{\Box^{-1}_\text{ret}} 
\bigl[r^\eta\,\overline{\tau}^{\mu\nu}\bigr] + \mathcal{H}^{\mu\nu}\,.
\end{equation}
The second term, $\mathcal{H}^{\mu\nu}$, is a homogeneous solution of the wave
equation and will be discussed later. The first term is a particular retarded
solution of the GFEE~\eqref{GFEE} when PN-expanded in the near zone. It is
defined from the retarded Green's function of the wave operator in $d+1$
space-time dimensions as~\cite{BBBFMc}
\begin{align}\label{retsol}
\overline{\Box^{-1}_\text{ret}} \bigl[r^\eta\,\overline{\tau}^{\mu\nu}\bigr] &
= - \frac{\tilde{k}}{4\pi} \int
\ud^d\mathbf{x}'
\,\vert\mathbf{x}'\vert^\eta \\ 
&\times\overline{\int_1^{+\infty} \ud z \,\gamma_{\frac{1-d}{2}}(z)\,
\frac{\overline{\tau}^{\mu\nu}(\mathbf{x}',t-z\vert\mathbf{x}-\mathbf{x}'\vert/c)}
       {\vert\mathbf{x}-\mathbf{x}'\vert^{d-2}}}\,,\nonumber
\end{align}
where the overbar refers to the PN expansion (see, notably, Appendix~A
in~\cite{BBBFMc}),
$\tilde{k}=\frac{\Gamma(\frac{d}{2}-1)}{\pi^{\frac{d}{2}-1}}$ ($\Gamma$ being
the Eulerian function) and
\begin{equation}\label{gamma}
\gamma_{\frac{1-d}{2}}(z) =
\frac{2\sqrt{\pi}}{\Gamma(\frac{3-d}{2})\Gamma(\frac{d}{2}-1)}
\,\big(z^2-1\bigr)^{\frac{1-d}{2}}\,,
\end{equation}
with the normalization condition $\int_1^{+\infty} \ud z\,\gamma_{\frac{1-d}{2}}(z) = 1$.

We have introduced in~\eqref{retsol} a factor $r^\eta$ multiplying the PN
source term. Such a factor is similar to the regulator $r^B$ entering the
general solution of the matching equation in 3 dimensions~\cite{B98mult, PB02,
  BFN05}. However, an important difference is that, here, we do not need to
take a ``finite part'' after integration (as we do in 3 dimensions). Indeed,
the regulator $r^\eta$ is inserted into the solution in $d=3+\varepsilon$
dimensions so that it acts ``on the top'' of dimensional regularization. Our
prescription is thus simply that we must consider first the limit $\eta\to 0$
for any generic dimension $d$ (i.e., avoiding integral values of $d$) and
check that, although divergences $\propto 1/\eta$ can occur in individual
terms, this limit is \textit{finite} for the sum of terms we consider. Only
afterwards do we apply the limit $\varepsilon\to 0$ and look for the presence
of poles $1/\varepsilon$. This regularization will be called the
``$\varepsilon\eta$'' regularization.

The contribution of the particular solution [i.e., the first term
in~\eqref{PNsol}] to the Fokker Lagrangian has been computed in Ref.~\cite{BBBFMa}.
The PN order to which one must truncate the metric to be inserted so as to
control the Lagrangian up to a given $n$PN order is determined by the method
``$n+2$'' (see Sec.~IV A in~\cite{BBBFMa}): Focusing on the conservative
dynamics, i.e., neglecting dissipative odd PN contributions, the
various metric components, in the guise
$\overline{h}=(\overline{h}^{00ii}, \overline{h}^{0i}, \overline{h}^{ij})$
with the notation
\begin{equation}\label{not00ii}
\overline{h}^{00ii} = 
\frac{2}{d-1}\left[(d-2)\overline{h}^{00}+\overline{h}^{ii}\right]\,,
\end{equation}
are to be inserted into the action up to the orders
$(c^{-n-2}, c^{-n-1},c^{-n-2})$ inclusively when $n$ is even, and up to the orders
$(c^{-n-1}, c^{-n-2},c^{-n-1})$ inclusively when $n$ is odd. At the 4PN order,
this means that the metric components are required up to the orders
$(c^{-6}, c^{-5},c^{-6})$. We parametrize the metric with the help of certain
potentials defined in $d$ dimensions; the most important are $V$, $V_i$ and
$\hat{W}_{ij}$, which enter at lowest order:
\begin{subequations}\label{metricpot}
\begin{align} 
\overline{h}^{00ii} &= -\frac{4}{c^{2}}V +
\calO\left(c^{-4}\right)\,,\\
\overline{h}^{0i} &= - \frac{4}{c^{3}} V_{i} +
\calO\left(c^{-5}\right)\,,\\
\overline{h}^{ij} &= - \frac{4}{c^{4}}\biggl(\hat{W}_{ij} -
\frac{1}{2} \delta_{ij} \hat{W}_{kk}\biggr) +
\calO\left(c^{-6}\right) \,.
\end{align}
\end{subequations}
[See Eq.~(4.14) in~\cite{BBBFMa} for the complete parametrization to the
desired accuracy $(c^{-6}, c^{-5},c^{-6})$.] The PN potentials obey a sequence
of iterated flat space-time wave equations in $d$ dimensions. Defining the
particles' mass, current and stress densities as
$\sigma = \frac{2}{d-1}[(d-2)T^{00} + T^{ii}]/c^2$, $\sigma_i = T^{0i}/c$, and
$\sigma_{ij} = T^{ij}$, we have
\begin{equation}\label{potVVi}
\Box V = - 4 \pi G\, \sigma\,,\qquad\Box V_{i} = - 4 \pi G\,\sigma_i\,,
\end{equation}
together with the more complicated nonlinear potential
\begin{equation}\label{potWij}
\!\!\!\!\Box \hat{W}_{ij} = -4\pi G\biggl(\sigma_{ij}
-\delta_{ij}\,\frac{\sigma_{kk}}{d-2}\biggr)
-\frac{d-1}{2(d-2)}\partial_i V \partial_j V\,.
\end{equation}

In the conservative dynamics, these potentials are generated by the
standard symmetric propagator; this corresponds to the first term in
Eq.~\eqref{PNsol}, with the retarded inverse d'Alembertian operator
$\Box^{-1}_\text{ret}$ replaced by the symmetric one. The resulting
conservative dynamics is characterized by an equal amount of incoming
and outgoing radiation. In the language of EFT, where the perturbative
expansion is achieved with Feynman diagrams, the conservative sector
is defined by diagrams that have no external graviton lines --- the
so-called ``radiative'' gravitons~\cite{FSrevue}. At the 4PN order, in
the conservative sector, a process appears in which the graviton is
emitted and then reabsorbed by the particles, and interacts with the
total particles' mass through a ``potential'' graviton. This is the
tail effect, which has been computed in the context of EFT in
Refs.~\cite{FStail,GLPR16}.

\section{The last ambiguity parameter} 
\label{sec:amb}

In our formalism, the computation of the last ambiguity parameter is achieved
by means of a consistent derivation of the tail effect at the 4PN order in $d$
dimensions, following the rules of the $\varepsilon\eta$ regularization. This
effect is described by the second term in Eq.~\eqref{PNsol}, which
--- as a consequence of the matching equation~\eqref{matching} --- is a specific
homogeneous solution of the wave equation, regular when $r\to 0$. Hence it is
of the form
\begin{equation}\label{Htail}
\mathcal{H}^{\mu\nu}(\mathbf{x},t) = \sum_{\ell=0}^{+\infty}
\sum_{j=0}^{+\infty} \frac{1}{c^{2j}}\,\Delta^{-j}\hat{x}_L \,f_L^{(2j)\mu\nu}(t)\,,
\end{equation}
where the superscript $(2j)$ refers to time derivatives, $L=i_1\cdots i_\ell$
is a multi-index made of $\ell$ spatial indices, $\hat{x}_L$ is the symmetric
trace-free (STF) product of $\ell$ spatial vectors $x^i$, and the $\ell$
summations on the dummy spatial indices $L$ are omitted. The $j$-th iterated
inverse Poisson operator $\Delta^{-j}$ acts on $\hat{x}_L$ as
\begin{equation}\label{poissonj}
  \Delta^{-j} \hat{x}_L 
= \frac{\Gamma(\ell+\frac{d}{2})}{\Gamma(\ell+j+\frac{d}{2})} \,
\frac{r^{2j} \hat{x}_L}{2^{2j} j!}\,.
\end{equation}

Most importantly, the function $f_L^{\mu\nu}(t)$ depends on the multipole
expansion $\mathcal{M}(\Lambda^{\mu\nu})$ of the gravitational source term in
the GFEE~\eqref{GFEE}. This reflects the fact that the PN-expanded solution in
the near zone is sensitive, via the matching equation~\eqref{matching}, to the
boundary conditions obeyed by the radiation field, in particular the
no-incoming radiation condition at past null infinity. We have shown that~\cite{BBBFMc}
\begin{align}\label{fLt}
f^{\mu\nu}_L(t) &= \frac{(-)^{\ell+1}\tilde{k}}{4\pi \ell!}
\int_1^{+\infty} \ud z \,\gamma_{\frac{1-d}{2}}(z) \\&\quad\times \int
\ud^d\mathbf{x}'\,\vert\mathbf{x}'\vert^\eta \,\hat{\partial}'_L\!
\left[\frac{\mathcal{M}(\Lambda^{\mu\nu})(\mathbf{y},t - z
    r'/c)}{{r'}^{d-2}}\right]_{\mathbf{y}=\mathbf{x}'}\,,\nonumber
\end{align}
where $\hat{\partial}'_L$ denotes the STF projection of a product of $\ell$
partial derivatives $\partial/\partial{x'}^i$, being understood that the
vector $y^i$ is to be treated as a constant when differentiating and replaced
by ${x'}^i$ only afterwards. Observe that Eq.~\eqref{fLt} is also
defined with the $\varepsilon\eta$ regularization.

In practice, the multipolar field $\mathcal{M}(h^{\mu\nu})$ is computed by
means of the so-called multipolar-post-Minkowskian (MPM) algorithm~\cite{BD86,
  B98mult},
\begin{equation}\label{MPMdef}
\mathcal{M}(h^{\mu\nu}) = h^{\mu\nu}_\text{MPM}\,.
\end{equation}
The MPM field represents the most general solution of the vacuum GFEE outside
the matter source. It consists of a formal post-Minkowskian (or post-linear)
expansion
\begin{equation}\label{MPMexp}
h^{\mu\nu}_\text{MPM} = \sum_{n=1}^{+\infty} G^n h^{\mu\nu}_{n}\,,
\end{equation}
with each post-Minkowskian coefficient $h^{\mu\nu}_{n}$ given in the form of a
multipole expansion. The MPM algorithm starts from the most general multipolar
solution of the linearized GFEE~\cite{Th80},
\begin{subequations}\label{h1can}\begin{align}
h^{00}_{1} &= - \frac{4G}{c^2} \sum_{\ell=0}^{+\infty}
\frac{(-)^\ell}{\ell!} \,\partial_L\tilde{I}_L\,,\\
h^{0i}_{1} &= \frac{4G}{c^3} \sum_{\ell=1}^{+\infty}
\frac{(-)^\ell}{\ell!}
\,\partial_{L-1}\tilde{I}_{iL-1}^{(1)}\,,\\
h^{ij}_{1} &= - \frac{4G}{c^4} \sum_{\ell=2}^{+\infty}
\frac{(-)^\ell}{\ell!}
\,\partial_{L-2}\tilde{I}^{(2)}_{ijL-2}\,,
\end{align}\end{subequations}
where the mass-type multipole moments are denoted $I_L(t)$ with, in
particular, for the monopole case $\ell=0$, $I=M$ representing the constant
ADM mass; moreover, the tilde over the moments means
\begin{equation}\label{ILtilde}
\tilde{I}_L(t,r) =
\frac{\tilde{k}}{r^{d-2}}\int_1^{+\infty} \ud
z\,\gamma_{\frac{1-d}{2}}(z)\,I_L(t-z r/c)\,,
\end{equation}
which, in the monopole case, reduces to $\tilde{M}(r)=\tilde{k} M r^{2-d}$.
For our purpose, we ignore the corresponding current-type multipole moments,
which could be defined in $d$ dimensions by means of ``mixed Young tableaux''
(see~\cite{BDEI05dr} for a discussion).

In order to determine the dominant tail effect at the 4PN order, we shall
consider the quadratic interaction between the ADM mass $M$ and the varying
mass quadrupole moment $I_{ij}(t)$. Thus, we shall focus on the source term of the
vacuum GFEE corresponding to that interaction: $M \times I_{ij}$. Now, the full
source term reads in general
\begin{equation}\label{sourceterm}
N_\text{MPM}^{\mu\nu}=\mathcal{M}(\Lambda^{\mu\nu}) 
= \sum_{n=2}^{+\infty} G^n N^{\mu\nu}_{n}\,,
\end{equation}
with
$\mathcal{M}(\Lambda^{\mu\nu})=\frac{16\pi G}{c^4}\mathcal{M}(\tau^{\mu\nu})$,
since the multipole expansion is a formal vacuum solution of the GFEE. Each
MPM coefficient in~\eqref{sourceterm} admits the decomposition
\begin{equation}\label{Nstruct}
N_{n}^{\mu\nu}(\mathbf{x},t) = \sum_{\ell=0}^{+\infty}\hat{n}_L N^{\mu\nu}_{nL}(r,t)\,,
\end{equation}
where $\hat{n}_L$ is the STF product of $\ell$ unit vectors $n^i=x^i/r$.
Plugging~\eqref{Nstruct} into~\eqref{fLt}, we obtain a related
post-Minkowskian expansion $f^{\mu\nu}_{L} = \sum_n G^n f^{\mu\nu}_{nL}$
with~\cite{BBBFMc}
\begin{align}\label{fLsimple}
f^{\mu\nu}_{nL} &= - \frac{1}{d+2\ell-2}
\,\int_1^{+\infty} \ud z
\,\gamma_{\frac{1-d}{2}-\ell}(z) \\&\qquad\quad\times\int_0^{+\infty} \ud
r'\,{r'}^{-\ell+1+\eta} \,N^{\mu\nu}_{nL}(r',t-z r'/c)\,.\nonumber
\end{align}
To order $n=2$, for the interaction $M \times I_{ij}$, the source term is a
sum of the type
\begin{equation}\label{Nstruct2}
N^{\mu\nu}_{2,L} = \sum {r}^{-k-2\varepsilon} \,\int_1^{+\infty} \ud y
\,y^p\,\gamma_{-1-\frac{\varepsilon}{2}}(y)\,F^{\mu\nu}_L(t-y r/c)\,,
\end{equation}
where the sum ranges over integers $k$, $p$, and the function $F^{\mu\nu}_L$
is made of the product of $M$ and components of $I_{ij}$; we have posed
$\varepsilon=d-3$. In that case, the expression~\eqref{fLsimple} becomes
\begin{align}\label{fLfinal}
f^{\mu\nu}_{2,L} &=
\sum \frac{(-)^{\ell+k}\,C_\ell^{p,k}}{2\ell+1+\varepsilon}
\,\frac{\Gamma(2\varepsilon-\eta)}{\Gamma(\ell+k-1+2\varepsilon-\eta)}
\nonumber\\ &\qquad\quad\times\int_0^{+\infty} \ud
\tau\,\tau^{-2\varepsilon+\eta}\,F_L^{(\ell+k-1)\mu\nu}(t-\tau)\,.
\end{align}
Interestingly, we could factorize out two of the three independent
integrations in~\eqref{fLsimple}--\eqref{Nstruct2} into a single (though
nontrivial looking) dimensionless coefficient
\begin{align}\label{coeffC}
C_\ell^{p,k} &= \int_1^{+\infty} \!\ud y
\,y^p\,\gamma_{-1-\frac{\varepsilon}{2}}(y)\\ &\qquad\times \int_1^{+\infty} \!\ud z
\,(y+z)^{\ell+k-2+2\varepsilon-\eta}\,\gamma_{-\ell-1-\frac{\varepsilon}{2}}(z)\,.\nonumber
\end{align}
The computation of this coefficient in analytic closed form is
described in the Appendix~D of~\cite{BBBFMc}.

We have applied the formulas~\eqref{fLfinal}--\eqref{coeffC} to obtain the
dominant tail effect in the metric at the 4PN order, which is given, according
to the matching procedure, by the homogeneous (regular at $r=0$)
solution~\eqref{Htail}. The result can be expressed in terms of a logarithmic kernel involving the combination
\begin{equation}\label{notL}
L(\tau) \equiv \ln\left(\frac{c\sqrt{\bar{q}}\,\tau}{2\ell_0}
  \right) - \frac{1}{2\varepsilon}\,,
\end{equation}
where $\bar{q} = 4\pi\,\ue^{\gamma_\text{E}}$, with $\gamma_\text{E}$ being
the Euler constant, and $\ell_0$ the dimensional regularization scale. Note
the appearance of a pole $\propto 1/\varepsilon$, which originates from the
lower integration bound $\tau\to 0$ in~\eqref{fLfinal} and is thus a UV pole.
Applying the latter precepts along with the $\varepsilon\eta$ regularization
and expanding the result at the 4PN order, we arrive at (with
$\mathcal{H}^{00ii} = \frac{2}{d-1}[(d-2)\mathcal{H}^{00}+\mathcal{H}^{ii}]$):
\begin{subequations}\label{Htailres}
\begin{align}
\mathcal{H}^{00ii} &= \frac{8 G_N^2M}{15 c^{10}} x^{ij}
\!\!\int_0^{+\infty}\!\!\!\ud\tau\biggl[ L(\tau) +
 \frac{61}{60}\biggr]I^{(7)}_{ij}(t-\tau) \nonumber\\ &+
\calO\left(c^{-12}\right)\,,\\
\mathcal{H}^{0i} &= - \frac{8 G_N^2M}{3 c^{9}} x^j
\!\!\int_0^{+\infty}\!\!\!\!\!\!\!\ud\tau\biggl[ L(\tau) +
  \frac{107}{120}\biggr]I^{(6)}_{ij}(t-\tau) \nonumber\\ &+
\calO\left(c^{-11}\right) \,,\\
\mathcal{H}^{ij} &= \frac{8 G_N^2M}{c^{8}}
\!\!\int_0^{+\infty}\!\!\!\ud\tau\biggl[ L(\tau) +
  \frac{4}{5}\biggr]I^{(5)}_{ij}(t-\tau) \nonumber\\ &+
\calO\left(c^{-10}\right) \,.
\end{align}
\end{subequations}
We have made the important verification that the homogeneous
solution~\eqref{Htailres} is divergenceless up to the required order, i.e.,
$\partial_\nu\mathcal{H}^{0\nu}=\calO(c^{-13})$ and
$\partial_\nu\mathcal{H}^{i\nu}=\calO(c^{-12})$. We have also verified that
the first term in~\eqref{PNsol} is separately divergenceless (using the
matching equation for the considered interaction $M \times I_{ij}$). Thus, the
complete PN solution satisfies the harmonic gauge condition up to that order:
$\partial_\nu\overline{h}^{\mu\nu}=0$.

Finally, we insert these results into the Fokker action in order to compute the tail
contribution therein. The quadratic interactions yield compact-support
expressions depending on the values of the homogeneous
solution~\eqref{Htailres} at the locations of the particles. However, a cubic
term with noncompact support also needs to be consistently included in the
action at the 4PN order, so that~\cite{BBBFMa}
\begin{align}\label{Stail0}
S_\text{F}^\text{tail} &= \sum_A \!m_A c^2 \!\int\!\ud t
\!\left[-\frac{1}{8}\mathcal{H}^{00ii}_A +
  \frac{1}{2c}\mathcal{H}^{0i}_A
  v^i_A \right. \nonumber \\ & \qquad 
\left. -\frac{1}{4c^2}\overline{\mathcal{H}}^{ij}_A
  v^i_Av^j_A\right] \nonumber\\ &
- \frac{1}{32\pi G}\frac{d-1}{d-2}\int\!\ud t\!\int
\ud^{d}\mathbf{x} \,\mathcal{H}^{ij}\partial_i V\partial_j
V \,.
\end{align}
This cubic term has a two-fold origin: it comes from (i) a direct cubic term
$\sim h\partial h\partial h$ in the action, and (ii) the quadratic
nonlinearity in the source of the potential $\hat{W}_{ij}$
[see Eq.~\eqref{potWij}]. Inserting Eqs.~\eqref{Htailres} into~\eqref{Stail0}, we
observe that the noncompact support piece elegantly combines with the other
terms to give a simple expression quadratic in the time derivatives of the
quadrupole moment $I_{ij}$. In the end, we get the tail contribution to the
action:
\begin{align}\label{Stail}
S_\text{F}^\text{tail} &= \frac{2G_N^2M}{5 c^8} 
\int_{-\infty}^{+\infty}\ud t\,I_{ij}^{(3)}(t) \\ 
&\qquad\quad\times\int_0^{+\infty} \ud\tau \biggl[
 L(\tau) + \frac{41}{60} \biggr] I_{ij}^{(4)}(t-\tau) \,,\nonumber
\end{align}
which can be rewritten in a manifestly time-symmetric way (under time
reversal) by means of a Hadamard partie finie (Pf) integral as
\begin{equation}\label{Stail2}
S_\text{F}^\text{tail} = \frac{G_N^2M}{5 c^8} \,\mathop{\text{Pf}}_{\tau_0}
\int\!\!\!\int \frac{\ud t\ud t'}{\vert t-t'\vert}
I_{ij}^{(3)}(t)\,I_{ij}^{(3)}(t')\,,
\end{equation}
where $\tau_0$ denotes the usual cut-off scale, here given by
$\tau_0 = \frac{2\ell_0}{c \sqrt{\bar{q}}}\,
\text{exp}[\frac{1}{2\varepsilon}-\frac{41}{60}]$.

Equations~\eqref{Stail}--\eqref{Stail2} describe the conservative part of the
tail effect at the 4PN order. It is shown in~\cite{BBBFMc} that, modulo an
unphysical shift of the particle's world-lines, the UV pole present
in~\eqref{Stail}--\eqref{Stail2} cancels out the corresponding IR pole
entering the gravitational part of the Fokker action computed with the method
$n+2$; furthermore, the associated dimensional regularization scale $\ell_0$
cleanly disappears from the final Lagrangian.

The result~\eqref{Stail}--\eqref{Stail2} closes our ambiguity-free derivation
of the 4PN equations of motion. Indeed, we found in~\cite{BBBFMc} that the
``last'' ambiguity parameter, say $\kappa$, which is equivalent to the
ambiguity parameter of the Hamiltonian formalism~\cite{DJS14}, is precisely
given by the numerical constant entering the tail term when evaluated in
$3+\varepsilon$ dimensions, beyond the pole $1/\varepsilon$. Now, the value we
obtain for this constant in~\eqref{Stail}, i.e., $\kappa = \frac{41}{60}$, is
in perfect agreement with that determined in~\cite{BBBFMb, BBBFMc} so as to
recover GSF calculations of the conserved energy and periastron advance for
circular orbits in the small mass-ratio limit.

Let us point out (as remarked in~\cite{BBBFMc}) that the latter value of
$\kappa$ is exactly the one found in the computation of the tail effect
through EFT methods (see Eq.~(3.3) in~\cite{GLPR16}). This confirms that the
EFT Lagrangian, when it is completed by all the instantaneous (nontail) terms
up to the 4PN order, will be ambiguity-free like ours, and in agreement with
GSF calculations.

We also want to stress the nice correspondence between the EFT approach and our
formalism. In the EFT, the tail effect is computed as a Feynman diagram with
one graviton emitted and absorbed by the particles, and one ``potential''
graviton responsible for the interaction with the total mass $M$. In our work,
the tail effect is the consequence of the second term in Eq.~\eqref{PNsol},
which represents a crucial additional homogeneous solution imposed by the
matching between the near and far zones. In this respect, it seems
that the lack of a consistent matching between the near and far zones in the
ADM Hamiltonian formalism~\cite{JaraS12, JaraS13, DJS14, DJS16}, i.e., an
analogue of our Eqs.~\eqref{matching}--\eqref{PNsol}, forces this formalism to
be still plagued by one ambiguity parameter (denoted $C$ in~\cite{DJS14}).

\acknowledgments

L.Be. acknowledges financial support provided under the European Union's H2020
ERC Consolidator Grant ``Matter and strong-field gravity: New frontiers in
Einstein's theory'' grant agreement no. MaGRaTh646597.

\appendix

\bibliography{ListeRef_MBBF}

\end{document}